\documentclass[a4paper,11pt]{article}
\pdfoutput=1 % if your are submitting a pdflatex (i.e. if you have
\usepackage{times}
\usepackage{physics}
\usepackage{jcappub} % for details on the use of the package, please
\usepackage{hyperref}
\usepackage[T1]{fontenc} % if needed
\usepackage{hyperref}
\usepackage{orcidlink}

\title{\boldmath Thermodynamic Geometry, Heat Engines, and Topology of Sharma–Mittal ModMax-dRGT Black Holes
}
\author[a,b,c]{Hassan Hassanabadi~\orcidlink{0000-0001-7487-6898}}
\author[d]{Abdullah Guvendi~\orcidlink{0000-0003-0564-9899}}
\author[e,f]{Dhruba Jyoti Gogoi~\orcidlink{0000-0002-4776-8506}}
\author[g]{Semra Gurtas Dogan~\orcidlink{0000-0001-7345-3287}}
\author[h]{Farokhnaz Hosseinifar~\orcidlink{0009-0003-7057-451X}}

\affiliation[a]{Department   of   Physics, Faculty of Science,   University   of   Hradec   Kr\'{a}lov\'{e},  Rokitansk\'{e}ho 62, 500   03   Hradec   Kr\'{a}lov\'{e},   Czechia}
\affiliation[b]{Khazar University, Department of Physics and Electronics, 41 Mahsati Str, AZ1096, Baku, Azerbaijan}
\affiliation[c]{Al-Farabi Kazakh National University, Al-Farabi ave. 71, 050040 Almaty, Kazakhstan}
\affiliation[d]{Department of Basic Sciences, Erzurum Technical University, 25050 Erzurum, Türkiye}
\affiliation[e]{Department of Physics, Madhabdev University, Narayanpur, Lakhimpur 784164, Assam, India}
\affiliation[f]{Jadara Research Center, Jadara University, Irbid 21110, Jordan}
\affiliation[g]{Department of Medical Imaging Techniques, Hakkari University, 30000, Hakkari, Türkiye}
\affiliation[h]{Center for Theoretical Physics, Khazar University, 41 Mehseti Street, Baku, AZ-1096, Azerbaijan}
\emailAdd{hha1349@gmail.com (Corresponding Author)}
\emailAdd{abdullah.guvendi@erzurum.edu.tr}
\emailAdd{moloydhruba@yahoo.in}
\emailAdd{semragurtasdogan@hakkari.edu.tr}
\emailAdd{f.hoseinifar94@gmail.com}

\abstract{
	We investigate the thermodynamic structure of charged AdS black holes in ModMax nonlinear electrodynamics coupled to dRGT-like massive gravity, incorporating Sharma--Mittal entropy corrections. The thermodynamic geometry is analyzed using the Weinhold metric in the parameter space spanned by the horizon radius and electric charge. The resulting thermodynamic Ricci scalar characterizes effective microscopic interactions, with curvature singularities signaling extremal boundaries and degeneracies of the thermodynamic metric. We further construct a rectangular black hole heat engine in the extended phase space and derive an exact expression for its efficiency, demonstrating how the ModMax parameter and massive-gravity couplings influence the enthalpy-based conversion of heat into work, while the Sharma--Mittal parameters modify the Carnot bound through corrections to the black-hole temperature. Finally, a topological analysis of the corrected temperature and generalized free energy reveals both conventional and novel critical points, and the associated conserved topological charge is investigated.
}

\begin{document}
	\maketitle
	\flushbottom
	%%%%%%%%%%%%%%%%%%%%%%%%%%%%%%%%%%%%%%%%%%%%%%%%%%%%
	%%%%%%%%%%%%%%%%%%%%%%%%%%%%%%%%%%%%%%%%%%%%%%%%%%%%
	\section{Introduction}\label{Sec1}
	%%%%%%%%%%%%%%%%%%%%%%%%%%%%%%%%%%%%%%%%%%%%%%%%%%%%
	%%%%%%%%%%%%%%%%%%%%%%%%%%%%%%%%%%%%%%%%%%%%%%%%%%%%
	Black hole thermodynamics has become one of the most effective theoretical tools for investigating the interface between gravitational dynamics,
    quantum effects, and statistical behavior \cite{davies1978thermodynamics,sucu2025charged,parikh2000hawking,bekenstein1980black,sorkin1998statistical,falls2014black,el2017quantum,alonso2023thermodynamics}. In addition to the conventional thermodynamic laws, geometric formulations provide a deeper insight into the internal structure of black hole systems. Thermodynamic geometry treats the space of equilibrium states as a geometric manifold, where the curvature reflects the correlation properties of the underlying microscopic degrees of freedom \cite{zulkowski2012geometry,brandner2020thermodynamic,sarkar2006thermodynamic,aaman2003geometry,wei2018thermodynamic}. In this picture, divergences in the thermodynamic curvature indicate special thermodynamic regimes, such as extremal configurations, critical phenomena, or degeneracies of the equilibrium description \cite{ruppeiner2012thermodynamic,fonda2019thermodynamic,ruppeiner2010thermodynamic}. Moreover, the sign of the curvature can provide qualitative information regarding whether the dominant
    effective microscopic interactions are attractive or repulsive \cite{wei2019repulsive}.
	
	This geometric perspective is especially valuable for understanding black holes that emerge from extended gravitational theories and nonlinear electromagnetic models \cite{kar2024novel,walia2024exploring,bokulic2024thermodynamics,capozziello2011extended}. In these cases, the thermodynamic behavior is influenced not only by the horizon radius, charge, and pressure, but also by additional parameters associated with nonlinear electrodynamics and massive gravity \cite{nam2018non,dehghani2021asymptotically,rehan2026thermodynamics}. Entropy corrections further enrich this structure by modifying the system's temperature and stability properties \cite{kruglov2021einstein}. For instance, the Sharma--Mittal entropy introduces a generalized nonextensive correction that modifies the thermal response of the black hole \cite{frank2002generalized}. Recent studies have further demonstrated that the Sharma--Mittal entropy represents a particular case of a more general entropy framework, which has been successfully applied to black hole thermodynamics and cosmology \cite{nojiri2022nonextensive,nojiri2022early}. Within this broader setting, the thermodynamic properties of black holes associated with generalized entropies, including their effects on Hawking temperature, stability, and phase transitions, have been systematically explored \cite{nojiri2021area,nojiri2026modified,Elizalde2025Universe}. Consequently, thermodynamic geometry provides a useful method for distinguishing the distinct effects of nonlinear electromagnetic fields, massive gravity interactions, and entropy deformations on the effective microscopic behavior of the horizon's degrees of freedom.
	
	A complementary perspective is provided by black hole heat engines within the extended phase-space formulation \cite{opatrny2012black,hendi2018black}. In this approach, the cosmological constant is treated as thermodynamic pressure, while the black hole mass plays the role of enthalpy \cite{panah2025black}. This allows black holes to be interpreted as working substances that perform mechanical work through cycles in the pressure-volume plane. The corresponding efficiency measures how efficiently absorbed heat is converted into useful work and therefore gives a direct quantitative probe of the underlying theory. For Sharma--Mittal corrected ModMax-dRGT black holes, the heat-engine framework is especially useful because different sectors affect the engine in distinct ways: the ModMax parameter changes the effective charge sector, massive-gravity couplings modify the enthalpy and heat input, and Sharma--Mittal corrections influence the corrected temperature and the Carnot bound. Accordingly, the joint study of thermodynamic geometry and heat-engine efficiency offers a coherent framework for analyzing both microscopic interaction patterns and macroscopic energy-conversion properties of these modified black holes.
	
	Over the past several years, the thermodynamic characterization of black holes, specifically regarding phase structures, stability, and critical phenomena, has evolved into an important point of gravitational physics. 
    Among the diverse analytical frameworks emerging in this field, the topological study of black hole thermodynamics has distinguished itself as a particularly insightful perspective. This pioneering formalism, initially proposed by Wei et al. \cite{wei2022topology}, is inspired by Duan’s $\phi-$mapping topological current theory \cite{duan1984structure}. By treating thermodynamic critical points as topological defects within a vector field, typically constructed from potentials such as the Hawking temperature, this method allows for the classification of critical points through the assignment of conserved topological charges \cite{wei2026topology}. This model has been extensively adopted to analyze a wide spectrum of configurations, ranging from modified gravity theories and black holes immersed in dark matter halos to rotating and higher-dimensional scenarios \cite{sadeghi12024thermodynamic,hosseinifar2026rotating,yerra2022topology,yerra22022topology,fan2023topological,shahzad2024topological}.
	
	Subsequent developments have extended this formalism beyond temperature-based vector fields to the generalized free energy landscape \cite{wei2022black}. This extension has enabled a deeper  exploration of black hole stability and global phase behavior. By applying this methodology to diverse gravitational systems, including various modified gravity theories and higher-dimensional black holes, researchers have successfully established accurate topological classification schemes \cite{wu2023topological,liu2023topological,brzo2025thermodynamic,wu12023topological,wu2024thermodynamical,wu2023consistent,liu2025thermodynamic,zhu2024topological,fang2023revisiting,hazarika2024thermodynamic,hassanabadi2026astrophysical}. Ultimately, this geometric framework serves as a robust analytical tool for elucidating phase transitions and the global stability of black holes within the extended phase space, independent of the specific dynamical details of the underlying theory.
	
	The remainder of the paper is organized as follows: In Section \ref{Sec2}, we introduce the metric of the black hole under investigation and derive both the Hawking and the modified temperatures. Section \ref{Sec3} is devoted to thermodynamic geometry, where we examine the thermodynamic curvature scalar and analyze its sign. 
    In Section \ref{Sec4}, we evaluate the efficiency of the heat engine utilizing a rectangular thermodynamic cycle. We then compute the heat input, compare the efficiency with the Maxwell and Carnot limits, and investigate the influence of massive gravity parameters on these processes. Section \ref{Sec5} focuses on the topological analysis of thermodynamic potentials. By identifying the critical points of the modified temperature and the generalized free energy, we determine the universal topological class of the black hole under the specified parameter constraints. Finally, Section \ref{Sec6} presents the conclusions of this work.

	\section{Black Hole Geometry and Thermodynamic Setup}\label{Sec2}
	
	Before proceeding to the thermodynamic geometry, heat-engine analysis, and topological characterization of the system, we briefly summarize the geometric background and the principal thermodynamic quantities associated with the Sharma--Mittal corrected ModMax-dRGT black hole. The thermodynamic properties of this solution have been investigated in detail in Ref.~\cite{panah2025black}; therefore, our aim in this section is not to revisit the complete thermodynamic analysis but rather to establish the notation and collect the essential relations that will be employed throughout the subsequent sections. We consider the static and spherically symmetric black hole solution introduced in Ref.~\cite{panah2025black}. The spacetime geometry is described by the line element
	\begin{equation}\label{ds2}
		ds^2=-f(r)dt^2+\frac{1}{f(r)}dr^2+r^2\left(d\theta^2+\sin^2\theta\, d\phi^2\right),
	\end{equation}
	where the metric function takes the form
	\begin{equation}
		f(r)=1-\frac{2M}{r}-\frac{\Lambda r^2}{3}
		+\frac{e^{-\gamma}q^2}{r^2}
		+C m_g^2\left(Cc_2+\frac{c_1r}{2}\right).
	\end{equation}
	This solution incorporates simultaneously the effects of a negative cosmological constant, ModMax nonlinear electrodynamics, and dRGT massive gravity. Here, $M$ denotes the gravitational mass of the black hole, $q$ is the electric charge, and $\Lambda$ represents the cosmological constant. The dimensionless parameter $\gamma$ controls the nonlinear electromagnetic contribution associated with the ModMax sector, while $m_g$ corresponds to the graviton mass. Furthermore, $C$ is a positive constant originating from the reference metric and $c_1$, $c_2$ are coupling parameters entering the massive-gravity potential~\cite{panah2025black}. The event horizon radius $r_+$ is determined by the largest positive root of the equation
	\begin{equation}
		f(r_+)=0.
	\end{equation}
	Since the horizon radius provides the natural thermodynamic variable characterizing the equilibrium configuration, it is convenient to express the mass parameter in terms of $r_+$. Adopting the extended phase-space interpretation in which the cosmological constant is identified with the thermodynamic pressure according to \cite{panah2025black,kubizvnak2015black,kastor2009enthalpy}
	\begin{equation}
		P=-\frac{\Lambda}{8\pi},
	\end{equation}
	one obtains the mass function
	\begin{equation}
		\label{eq:mass_therm_geom}
		M=
		\frac{1}{12}
		r_+
		\Big(
		6C^2c_2m_g^2
		+
		3Cc_1m_g^2r_+
		+
		16\pi Pr_+^2
		+
		6
		\Big)
		+
		\frac{e^{-\gamma}q^2}{2r_+}.
	\end{equation}
	Within the extended thermodynamic framework, the mass should be interpreted as the enthalpy of the black hole spacetime. Equation (\ref{eq:mass_therm_geom}) explicitly demonstrates how the AdS pressure, the nonlinear electromagnetic sector, and the massive-gravity contributions jointly modify the energetic content of the horizon. The Bekenstein-Hawking entropy associated with the event horizon obeys the standard area law, \(\mathcal{S}=\pi r_+^2\). Using Eq.~(\ref{eq:mass_therm_geom}), the corresponding Hawking temperature can be obtained through \cite{hawking1975particle,bekenstein2020black}
	\begin{equation}
		T_{\rm H}
		=
		\frac{\partial M}{\partial \mathcal{S}},
	\end{equation}
	which yields
	\begin{equation}
		\begin{aligned}
			T_{\rm H}
			=
			2Pr_+
			+
			\frac{e^{-\gamma}}
			{4\pi r_+^3}
			\Big(
			e^\gamma C^2c_2m_g^2r_+^2
			+
			e^\gamma Cc_1m_g^2r_+^3
			-
			q^2
			+
			e^\gamma r_+^2
			\Big).
		\end{aligned}
	\end{equation}
	As discussed extensively in Ref.~\cite{panah2025black}, this temperature function governs the local thermodynamic behavior of the black hole and encodes the competing effects of the electromagnetic, gravitational, and AdS sectors. To incorporate nonextensive entropy effects, we employ the Sharma--Mittal entropy formalism. The generalized entropy is given by \cite{sharma1975new,masi2005step}
	\begin{equation}
		\label{eq:SSM}
		\mathcal{S}_{\rm SM}
		=
		\frac{1}{R}
		\Big[
		(1+\delta\mathcal{S})^{R/\delta}
		-
		1
		\Big],
	\end{equation}
	where $R$ and $\delta$ denote the Sharma--Mittal deformation parameters. This generalized entropy provides a unified framework encompassing several nonextensive entropy models and introduces additional corrections to the thermodynamic description of the horizon. The corresponding modified temperature is obtained through
	\begin{equation}
		T_{\rm SM}=\frac{\partial M}{\partial \mathcal{S}_{\rm SM}}.
	\end{equation}
	Since both $M$ and $\mathcal{S}_{\rm SM}$ depend on the horizon radius $r_+$, the temperature can be rewritten in terms of derivatives with respect to $r_+$ as
	\begin{equation}
		T_{\rm SM}=\frac{dM/dr_+}{d\mathcal{S}_{\rm SM}/dr_+}.
	\end{equation}
	The entropy function in \eqref{eq:SSM} yields
	\begin{equation}
		\frac{d\mathcal{S}_{\rm SM}}{dr_+}=2\pi r_+\left(1+\pi \delta r_+^2\right)^{\frac{R}{\delta}-1},
	\end{equation}
	where the overall structure is fully determined by the nonextensive deformation parameters $R$ and $\delta$. For the mass function \eqref{eq:mass_therm_geom}, one obtains
	\begin{equation}
		\frac{dM}{dr_+}
		=
		\frac{1}{2}
		+4\pi P r_+^2
		+\frac{C^2 c_2 m_g^2}{2}
		+\frac{C c_1 m_g^2 r_+}{2}
		-\frac{e^{-\gamma}q^2}{2r_+^2}.
	\end{equation}
	Substituting these results gives
	\begin{equation}
		T_{\rm SM}=\frac{\frac{1}{2}+4\pi P r_+^2+\frac{C^2 c_2 m_g^2}{2}+\frac{C c_1 m_g^2 r_+}{2}-\frac{e^{-\gamma}q^2}{2r_+^2}}{2\pi r_+
			\left(1+\pi \delta r_+^2\right)^{\frac{R}{\delta}-1}}.
	\end{equation}
	Rearranging into a unified thermodynamic structure leads to
	\begin{equation}
		\begin{aligned}
			T_{\rm SM}=\left(1+\pi \delta r_+^2\right)^{1-\frac{R}{\delta}}\frac{
				C m_g^2 r_+^2 (C c_2 + c_1 r_+)
				+ 8\pi P r_+^4
				- e^{-\gamma} q^2
				+ r_+^2
			}{
				4\pi r_+^3
			}.
		\end{aligned}
	\end{equation}
	The resulting expression shows that the thermodynamic modification induced by the Sharma--Mittal structure enters exclusively through a multiplicative deformation factor originating from the entropy sector, while the geometric contributions from mass, charge, pressure, and massive gravity remain structurally unchanged. This leads to the compact representation
	\begin{equation}
		T_{\rm SM}=T_{\rm H}
		\left(1+\pi \delta r_+^2\right)^{1-\frac{R}{\delta}} ,   
	\end{equation}
	highlighting that the nonextensive correction acts as a global rescaling of the Hawking temperature without altering its underlying horizon dependence. Since the thermodynamic properties of the underlying ModMax--dRGT black hole have been extensively discussed in Ref.~\cite{panah2025black}, we shall not revisit stability conditions or phase structure here. Instead, $T_{\rm SM}$ will serve as the fundamental thermodynamic quantity for constructing thermodynamic geometry, evaluating curvature invariants, analyzing heat-engine efficiency, and classifying topological structures of equilibrium configurations in the subsequent sections.
	
	%%%%%%%%%%%%%%%%%%%%%%%%%%%%%%%%%%%%%%%%%%%%%%%%%%%%
	%%%%%%%%%%%%%%%%%%%%%%%%%%%%%%%%%%%%%%%%%%%%%%%%%%%%
	\section{Thermodynamic Geometry and Microscopic Structure}\label{Sec3}
	%%%%%%%%%%%%%%%%%%%%%%%%%%%%%%%%%%%%%%%%%%%%%%%%%%%%
	%%%%%%%%%%%%%%%%%%%%%%%%%%%%%%%%%%%%%%%%%%%%%%%%%%%%
	In order to gain further insight into the microscopic properties of the
	Sharma--Mittal corrected ModMax-dRGT black hole, we investigate its
	thermodynamic geometry. Thermodynamic geometry provides an effective
	macroscopic description of the statistical correlations associated with the
	underlying horizon degrees of freedom. In particular, divergences of the
	thermodynamic curvature are frequently associated with critical phenomena and
	phase transitions, while its sign provides qualitative information about the
	nature of the effective microscopic correlations.
	
	To avoid confusion, we denote the Ricci scalar associated with the spacetime
	metric by $R_{\rm sp}$, while the Ricci scalar associated with the
	thermodynamic metric is denoted by $\mathcal{R}_{\rm th}$. The symbol $R$
	continues to denote the Sharma--Mittal entropy parameter.
	
	For fixed pressure $P$, we choose the thermodynamic coordinates
	\begin{equation}
		X^i=(r_+,q),
	\end{equation}
	and define the thermodynamic metric through the Weinhold construction \cite{quevedo2008geometrothermodynamics,guo2022weinhold},
	\begin{equation}
		g^{\rm th}_{ij}
		=
		\frac{1}{T_{\rm SM}}
		\frac{\partial^2 M}{\partial X^i\partial X^j}.
		\label{eq:weinhold_metric}
	\end{equation}
	Here $M$ is the black hole mass and $T_{\rm SM}$ is the Sharma--Mittal
	corrected temperature.
	
	Using Eq. \eqref{eq:mass_therm_geom} one obtains
	\begin{align}
		\frac{\partial^2M}{\partial r_+^2}
		&=
		8\pi P r_+
		+
		\frac{q^2e^{-\gamma}}{r_+^3}
		+
		\frac{m_g^2Cc_1}{2},
		\\
		\frac{\partial^2M}{\partial r_+\partial q}
		&=
		-\frac{qe^{-\gamma}}{r_+^2},
		\\
		\frac{\partial^2M}{\partial q^2}
		&=
		\frac{e^{-\gamma}}{r_+}.
	\end{align}
	Therefore,
	\begin{equation}
		g^{\rm th}_{ij}
		=
		\frac{1}{T_{\rm SM}}
		\begin{pmatrix}
			8\pi P r_+
			+\dfrac{q^2e^{-\gamma}}{r_+^3}
			+\dfrac{m_g^2Cc_1}{2}
			&
			-\dfrac{qe^{-\gamma}}{r_+^2}
			\\[3mm]
			-\dfrac{qe^{-\gamma}}{r_+^2}
			&
			\dfrac{e^{-\gamma}}{r_+}
		\end{pmatrix}.
		\label{eq:metric_therm_geom}
	\end{equation}
	The determinant of the metric is
	\begin{equation}
		g_{\rm th}=
		\frac{e^{-\gamma}}{T_{\rm SM}^{2}}
		\left(
		8\pi P+\frac{m_g^2Cc_1}{2r_+}
		\right) .
		\label{eq:det_therm_geom}
	\end{equation}
	
	For compactness, we now introduce the following abbreviations:
	\begin{equation}
		\begin{aligned}
			a&\equiv e^{-\gamma},\qquad
			B\equiv 8\pi P,
			\qquad
			\lambda\equiv m_g^2Cc_1,
			\qquad
			\sigma\equiv m_g^2C^2c_2,
			\\
			H(r_+)&\equiv B+\frac{\lambda}{2r_+},
			\qquad
			\Delta(r_+,q)\equiv
			Br_+^4+
			\lambda r_+^3+
			(1+\sigma)r_+^2
			-aq^2,
			\\
			\Xi(r_+)&\equiv
			\left(1+\delta\pi r_+^2\right)^{1-R/\delta}.
		\end{aligned}
	\end{equation}
	With these definitions, the corrected temperature can be written as
	\begin{equation}
		T_{\rm SM}
		=
		\frac{\Delta(r_+,q)}{4\pi r_+^3}\,\Xi(r_+).
		\label{eq:T_compact}
	\end{equation}
	The metric in Eq.~\eqref{eq:metric_therm_geom} is conformally related to the
	Weinhold Hessian metric $h_{ij}$ as
	\begin{equation}
		g^{\rm th}_{ij}=\frac{1}{T_{\rm SM}}h_{ij}.
	\end{equation}
	According to Eq. \eqref{eq:weinhold_metric} and the above equation, $h_{ij}$ becomes $h_{ij}=\partial^2 M/(\partial X^i\partial X^j)$. The $d-$dimensional expression for $R_{\rm th}$ is given by 
	\begin{equation}
		\mathcal{R}_{\rm th} =\Omega^{-2}\left[R_h -2(d-1)\nabla^2 \ln \Omega- (d-1)(d-2)(\nabla\ln\Omega)\right],
	\end{equation}
	where $R_h$ is the Ricci scalar calculated from the Hessian metric $h_{ij}$.
	In two dimensions, the scalar curvature under this conformal transformation is
	\begin{equation}
		\mathcal{R}_{\rm th}
		=
		T_{\rm SM}
		\left[
		R_h+\nabla_h^2\ln T_{\rm SM}
		\right],
		\label{eq:conformal_curvature}
	\end{equation}
	where $R_h$ and $\nabla_h^2$ are calculated using the Hessian metric
	$h_{ij}$. Direct calculation gives
	\begin{equation}
		R_h=
		\frac{4Br_+ +\lambda}
		{r_+^2(2Br_+ +\lambda)^2}.
		\label{eq:Rh}
	\end{equation}
	Furthermore, defining
	\begin{equation}
		\Phi(r_+,q)
		\equiv \ln T_{\rm SM}
		=
		\ln \Delta(r_+,q)+\ln \Xi(r_+)-3\ln r_+ + {\rm const.},
	\end{equation}
	one obtains
	\begin{align}
		\partial_{r_+}\Phi
		&=
		\frac{4Br_+^3+3\lambda r_+^2+2(1+\sigma)r_+}{\Delta}
		+
		\frac{2\pi\delta r_+(1-R/\delta)}{1+\pi\delta r_+^2}
		-
		\frac{3}{r_+},
		\\
		\partial_q\Phi
		&=
		-\frac{2aq}{\Delta}.
	\end{align}
	The inverse Hessian metric is
	\begin{equation}\label{3.17}
		\begin{aligned}
			h^{r_+r_+}&=\frac{1}{r_+H},
			\\
			h^{r_+q}&=\frac{q}{r_+^2H},
			\\
			h^{qq}&=\frac{r_+}{a}+\frac{q^2}{r_+^3H},
		\end{aligned}
	\end{equation}
	and $\sqrt{h}=\sqrt{aH}$. Hence the general form of Hessian Laplacian appearing in
	Eq.~\eqref{eq:conformal_curvature} is
	\begin{equation}
		\nabla^2_h\Phi=\frac{1}{\sqrt{h}}\partial_i \left(\sqrt{h}h^{ij}\partial_j \Phi\right),\qquad\Phi=\ln T_{\rm SM}.
	\end{equation}
	By applying Eq. \eqref{3.17}, this expression reduces to
	\begin{align}
		\nabla_h^2\Phi
		=&
		\frac{1}{\sqrt{aH}}
		\partial_{r_+}
		\left[
		\sqrt{aH}
		\left(
		\frac{1}{r_+H}\partial_{r_+}\Phi
		+
		\frac{q}{r_+^2H}\partial_q\Phi
		\right)
		\right]
		\nonumber\\
		&+
		\frac{1}{\sqrt{aH}}
		\partial_q
		\left[
		\sqrt{aH}
		\left(
		\frac{q}{r_+^2H}\partial_{r_+}\Phi
		+
		\left(
		\frac{r_+}{a}+\frac{q^2}{r_+^3H}
		\right)
		\partial_q\Phi
		\right)
		\right].
		\label{eq:laplace_explicit}
	\end{align}
	Combining Eqs.~\eqref{eq:conformal_curvature}--\eqref{eq:laplace_explicit},
	the explicit thermodynamic Ricci scalar is
	\begin{equation}
			\mathcal{R}_{\rm th}
			=
			T_{\rm SM}
			\left\{
			\frac{4Br_+ +\lambda}
			{r_+^2(2Br_+ +\lambda)^2}
			+
			\nabla_h^2\ln T_{\rm SM}
			\right\}.
		\label{eq:Rth_explicit}
	\end{equation}
	This form is exact and is more useful than a fully expanded expression, because
	it displays separately the contribution from the Weinhold Hessian curvature and
	the conformal temperature factor generated by the Sharma--Mittal modification.
	
	The thermodynamic curvature diverges whenever the determinant of the metric
	vanishes,
	\begin{equation}
		8\pi P+\frac{m_g^2Cc_1}{2r_+}=0,
		\label{eq:det_zero_condition}
	\end{equation}
	or whenever $T_{\rm SM}=0$, which corresponds to the extremal boundary of the
	black hole configuration. These singularities indicate special points of the
	thermodynamic manifold where the ordinary equilibrium description becomes
	singular.
	%%%%%%%%%%%%%%%%%%%%%%%%%%%%%%%%%%%%%%%%%%%%%%%%%%%%
	%%%%%%%%%%%%%%%%%%%%%%%%%%%%%%%%%%%%%%%%%%%%%%%%%%%%
	\subsection{Sign of the Thermodynamic Curvature}
	%%%%%%%%%%%%%%%%%%%%%%%%%%%%%%%%%%%%%%%%%%%%%%%%%%%%
	%%%%%%%%%%%%%%%%%%%%%%%%%%%%%%%%%%%%%%%%%%%%%%%%%%%%
	The sign of $\mathcal{R}_{\rm th}$ provides useful qualitative information
	about the effective microscopic interaction structure associated with the
	horizon degrees of freedom. Since the fundamental microscopic constituents of
	the black hole are not known, the interpretation should be understood in terms
	of statistical correlations among effective horizon microstates rather than
	ordinary molecular interactions.
	
	Negative values of $\mathcal{R}_{\rm th}$ are generally interpreted as
	indicating predominantly attractive statistical correlations, whereas positive
	values are associated with predominantly repulsive correlations among the
	effective horizon microstates. For sufficiently small horizon radii, the charge
	sector dominates the thermodynamic response. Since the charge contribution
	enters through
	\begin{equation}
		q_{\rm eff}^{2}=q^2e^{-\gamma},
	\end{equation}
	the small-radius region is typically driven toward
	\begin{equation}
		\mathcal{R}_{\rm th}<0,
	\end{equation}
	indicating attractive microscopic correlations.
	
	For larger horizon radii, the electromagnetic contribution becomes subdominant
	and the massive-gravity sector becomes increasingly important. In particular,
	positive values of $c_1$ tend to push the curvature toward positive values,
	\begin{equation}
		\mathcal{R}_{\rm th}>0,
	\end{equation}
	which indicates the emergence of predominantly repulsive correlations. On the
	other hand, sufficiently negative values of $c_1$ may keep the curvature
	negative over a wider range of horizon radii.
	
	The ModMax parameter $\gamma$ plays a significant role in this interpretation.
	Increasing $\gamma$ suppresses the effective charge contribution through
	$e^{-\gamma}$; consequently, the attractive charge-dominated contribution is
	weakened and the thermodynamic curvature is generally shifted toward positive
	values. Thus ModMax nonlinearities can move the system from an
	attractive-correlation regime toward a repulsive-correlation regime.
	
	A particularly interesting point occurs when
	\begin{equation}
		\mathcal{R}_{\rm th}=0.
	\end{equation}
	The corresponding value of $r_+$ defines a crossover scale separating regions
	dominated by attractive and repulsive statistical correlations. This crossover
	can be interpreted as a change in the effective microscopic interaction regime
	of the black hole horizon fluid.
	
	Consequently, thermodynamic geometry provides a complementary description of
	the phase structure obtained from the Gibbs free-energy analysis. While the
	Gibbs free energy characterizes global thermodynamic stability, the
	thermodynamic curvature probes the effective microscopic interaction structure
	of the horizon degrees of freedom. Together, these approaches provide a more
	complete thermodynamic picture of the Sharma--Mittal corrected ModMax-dRGT
	black hole.
	%%%%%%%%%%%%%%%%%%%%%%%%%%%%%%%%%%%%%%%%%%%%%%%%%%%%
	%%%%%%%%%%%%%%%%%%%%%%%%%%%%%%%%%%%%%%%%%%%%%%%%%%%%
	\section{Heat Engine Efficiency in the Extended Phase Space}\label{Sec4}
	%%%%%%%%%%%%%%%%%%%%%%%%%%%%%%%%%%%%%%%%%%%%%%%%%%%%
	%%%%%%%%%%%%%%%%%%%%%%%%%%%%%%%%%%%%%%%%%%%%%%%%%%%%
	In the extended phase-space formulation of black hole thermodynamics, the cosmological constant is interpreted as a thermodynamic pressure,
	\begin{equation}
		P=-\frac{\Lambda}{8\pi},
	\end{equation}
	and the black hole mass is identified with the enthalpy of the gravitational system,
	\begin{equation}
		M\equiv H.
	\end{equation}
	The first law therefore takes the form \cite{cvetivc2011black}
	\begin{equation}
		dM=T_{\rm SM}dS_{\rm SM}+VdP+\Phi dq+\cdots ,
	\end{equation}
	where the thermodynamic volume conjugate to the pressure is
	\begin{equation}
		V=\left(\frac{\partial M}{\partial P}\right)_{S,q,c_i}
		=\frac{4\pi r_+^3}{3}.
		\label{eq:heatengine_volume}
	\end{equation}
	The existence of the nontrivial $P$-$V$ sector allows the black hole to be regarded as a thermodynamic working substance. In this picture, the work output is not ordinary mechanical motion of the horizon, but rather energy exchange associated with variations of the thermodynamic volume and the AdS vacuum pressure.
	%%%%%%%%%%%%%%%%%%%%%%%%%%%%%%%%%%%%%%%%%%%%%%%%%%%%
	%%%%%%%%%%%%%%%%%%%%%%%%%%%%%%%%%%%%%%%%%%%%%%%%%%%%
	%\subsection{Rectangular Heat-Engine Cycle}
	%%%%%%%%%%%%%%%%%%%%%%%%%%%%%%%%%%%%%%%%%%%%%%%%%%%%
	%%%%%%%%%%%%%%%%%%%%%%%%%%%%%%%%%%%%%%%%%%%%%%%%%%%%
	
\begin{figure}[htbp]
		\centering
		\includegraphics[width=6.2cm]{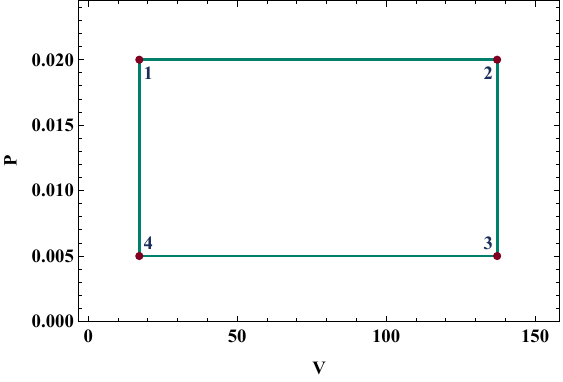} \hspace{0.2cm}
		\caption{Rectangular heat-engine cycle in the $P$-$V$ plane. The upper isobar corresponds to heat absorption, while the lower isobar corresponds to heat release.}
		\label{fig:pv_cycle_heat_engine}
	\end{figure}

	In Figure \ref{fig:pv_cycle_heat_engine}, we consider a rectangular cycle in the $P$-$V$ plane, consisting of two isobaric and two isochoric processes,
	\begin{equation}
		1\rightarrow2\rightarrow3\rightarrow4\rightarrow1,
	\end{equation}
	with corners
	\begin{equation}
		(P_1,V_1),\qquad (P_1,V_2),\qquad (P_4,V_2),\qquad (P_4,V_1),
	\end{equation}
	where
	\begin{equation}
		P_1>P_4,\qquad V_2>V_1.
	\end{equation}
	Using Eq.~\eqref{eq:heatengine_volume}, the two volumes are written as
	\begin{equation}
		V_i=\frac{4\pi r_i^3}{3},\qquad i=1,2.
	\end{equation}
	The mechanical work performed over one complete cycle is \cite{johnson2016born}
	\begin{equation}
		W=\oint P\,dV.
	\end{equation}
	For the rectangular cycle, this gives
	\begin{equation}
		W=(P_1-P_4)(V_2-V_1)
		=\frac{4\pi}{3}(P_1-P_4)\left(r_2^3-r_1^3\right).
		\label{eq:heatengine_work}
	\end{equation}
	
	%%%%%%%%%%%%%%%%%%%%%%%%%%%%%%%%%%%%%%%%%%%%%%%%%%%%
	%%%%%%%%%%%%%%%%%%%%%%%%%%%%%%%%%%%%%%%%%%%%%%%%%%%%
	%\subsection{Heat Input}
	%%%%%%%%%%%%%%%%%%%%%%%%%%%%%%%%%%%%%%%%%%%%%%%%%%%%
	%%%%%%%%%%%%%%%%%%%%%%%%%%%%%%%%%%%%%%%%%%%%%%%%%%%%
	
	For static black holes, both entropy and thermodynamic volume are functions of the horizon radius only. Therefore, along an isochoric branch $r_+$ remains constant and one has \cite{dolan2012pdv}
	\begin{equation}
		C_V=0.
	\end{equation}
	As a result, no heat is exchanged along the isochoric branches. The heat input is determined entirely by the upper isobar,
	\begin{equation}
		Q_H=M(P_1,r_2)-M(P_1,r_1).
	\end{equation}
	For the ModMax-dRGT black hole, the enthalpy is
	\begin{equation}
		M(r_+,P)=
		\frac{r_+}{2}
		+\frac{4\pi P r_+^3}{3}
		+\frac{q^2e^{-\gamma}}{2r_+}
		+\frac{m_g^2Cr_+}{2}
		\left(
		\frac{c_1r_+}{2}+c_2C
		\right).
		\label{eq:heatengine_enthalpy}
	\end{equation}
	Hence the heat absorbed along the upper isobar becomes
	\begin{align}
		Q_H
		=&
		\frac{r_2-r_1}{2}
		+\frac{4\pi P_1}{3}\left(r_2^3-r_1^3\right)
		+\frac{q^2e^{-\gamma}}{2}
		\left(\frac{1}{r_2}-\frac{1}{r_1}\right)
		\nonumber\\
		&+
		\frac{m_g^2Cc_1}{4}\left(r_2^2-r_1^2\right)
		+\frac{m_g^2C^2c_2}{2}(r_2-r_1).
		\label{eq:heatengine_QH}
	\end{align}
	Similarly, the heat released along the lower isobar is
	\begin{equation}
		Q_C=M(P_4,r_2)-M(P_4,r_1).
	\end{equation}

\begin{figure}[htbp]
		\centering
		\includegraphics[width=0.45\linewidth]{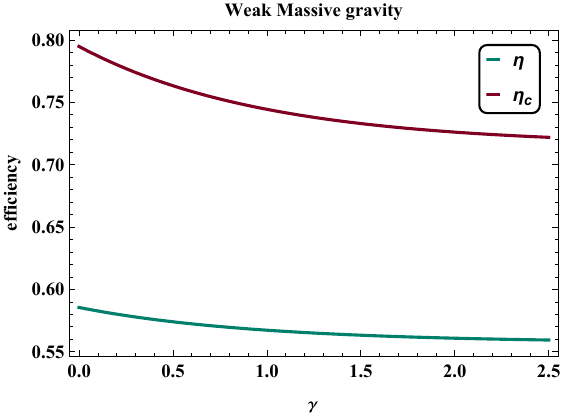}
		\includegraphics[width=0.45\linewidth]{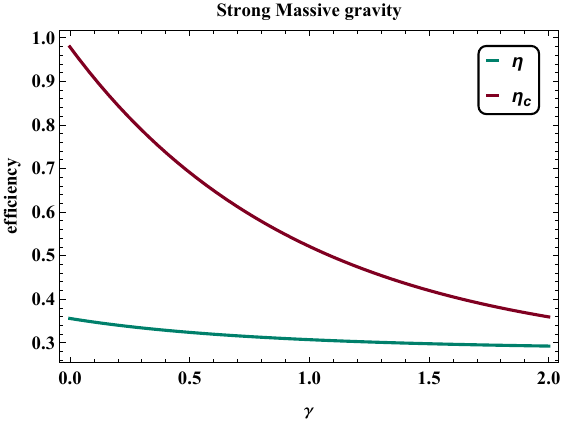}
		\caption{Heat-engine efficiency and Carnot bound as functions of the ModMax parameter considering $q=1$, $R=0.05$, and $\delta=0.1$ for both weak ($m_g=0.4,\,C=0.1,\,c_1=1,\,c_2=0.2$) and strong ($m_g=0.84,\,C=0.4,\,c_1=-10,\,c_2=23.0$) massive gravity cases. The point at $\gamma=0$ corresponds to the Maxwell limit.}
		\label{fig:eta_gamma}
	\end{figure}

	The heat-engine efficiency is defined by
	\begin{equation}
		\eta=\frac{W}{Q_H}.
	\end{equation}
	Using Eqs.~\eqref{eq:heatengine_work} and \eqref{eq:heatengine_QH}, the exact efficiency of the rectangular cycle is
	\begin{equation}
		\eta=
		\frac{
			\frac{4\pi}{3}(P_1-P_4)(r_2^3-r_1^3)
		}{
			\frac{r_2-r_1}{2}
			+\frac{4\pi P_1}{3}(r_2^3-r_1^3)
			+\frac{q^2e^{-\gamma}}{2}
			\left(\frac{1}{r_2}-\frac{1}{r_1}\right)
			+\frac{m_g^2Cc_1}{4}(r_2^2-r_1^2)
			+\frac{m_g^2C^2c_2}{2}(r_2-r_1)
		}.
		\label{eq:heatengine_eta}
	\end{equation}
	The Maxwell limit is obtained by setting
	\begin{equation}
		\gamma\rightarrow0,
		\qquad
		q^2e^{-\gamma}\rightarrow q^2.
	\end{equation}
	Therefore,
	\begin{equation}
		\eta_{\rm Maxwell}=\eta\big|_{\gamma=0}.
	\end{equation}
	For the representative cycle used in Figure~\ref{fig:eta_gamma}, one obtains $\eta_{\rm Maxwell}\simeq 0.5465$ and $\eta(\gamma=0.8)\simeq 0.5316$. These values are illustrative and depend on the chosen cycle parameters.

	%%%%%%%%%%%%%%%%%%%%%%%%%%%%%%%%%%%%%%%%%%%%%%%%%%%%
	%%%%%%%%%%%%%%%%%%%%%%%%%%%%%%%%%%%%%%%%%%%%%%%%%%%%
	\subsection{Sharma--Mittal Corrected Carnot Efficiency}
\begin{figure}[htbp]
		\centering
		\includegraphics[width=0.45\linewidth]{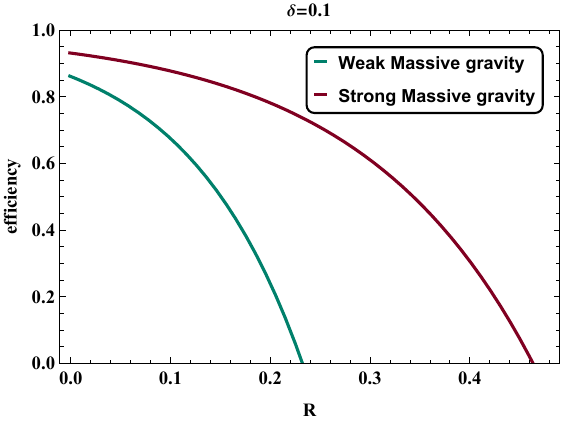}
		\includegraphics[width=0.45\linewidth]{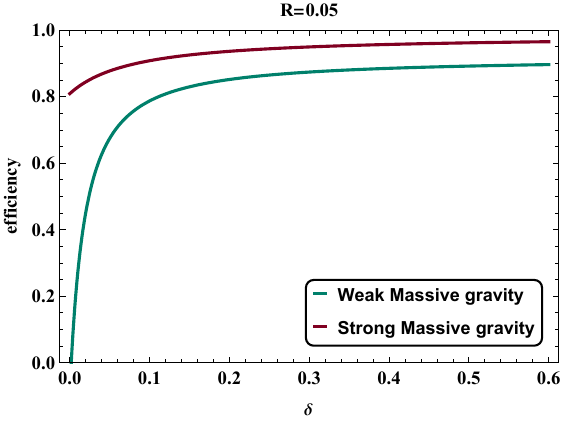}
		\caption{Effect of the Sharma--Mittal parameters on the Carnot efficiency where the electric charge is set to $q=1$, for weak ($m_g=0.4,\,C=0.1,\,c_1=1,\,c_2=0.2$) and strong ($m_g=0.8,\,C=0.1,\,c_1=-10,\,c_2=23.0$) massive gravity cases.
		For a fixed rectangular cycle, the mechanical efficiency is independent of $(R,\delta)$, while the Carnot bound changes through the corrected temperature.}
		\label{fig:carnot_sharma_mittal}
	\end{figure}

	%%%%%%%%%%%%%%%%%%%%%%%%%%%%%%%%%%%%%%%%%%%%%%%%%%%%
	%%%%%%%%%%%%%%%%%%%%%%%%%%%%%%%%%%%%%%%%%%%%%%%%%%%%
	The maximum theoretical efficiency is bounded by the Carnot efficiency,
	\begin{equation}
		\eta_C=1-\frac{T_C}{T_H}.
	\end{equation}
	For the rectangular cycle, we identify
	\begin{equation}
		T_H=T_{\rm SM}(P_1,r_2),
		\qquad
		T_C=T_{\rm SM}(P_4,r_1).
	\end{equation}
	The Sharma--Mittal corrected temperature is
	\begin{equation}
		T_{\rm SM}
		=
		T_0
		\left(1+\delta\pi r_+^2\right)^{1-R/\delta},
	\end{equation}
	where
	\begin{equation}
		T_0=
		2Pr_+
		+\frac{1}{4\pi}
		\left(
		\frac{1}{r_+}-\frac{q^2e^{-\gamma}}{r_+^3}
		\right)
		+\frac{m_g^2C}{4\pi}
		\left(
		c_1+\frac{c_2C}{r_+}
		\right).
	\end{equation}
	Thus the Sharma--Mittal corrected Carnot bound becomes
	\begin{equation}
		\eta_C
		=
		1-
		\frac{
			T_0(P_4,r_1)
			\left(1+\delta\pi r_1^2\right)^{1-R/\delta}
		}{
			T_0(P_1,r_2)
			\left(1+\delta\pi r_2^2\right)^{1-R/\delta}
		}.
		\label{eq:heatengine_carnot}
	\end{equation}
	The dependence of the Carnot bound on the Sharma–Mittal parameters $R$ and $\delta$ is shown in Figure \ref{fig:carnot_sharma_mittal}.
	Unlike the enthalpy-based efficiency in Eq.~\eqref{eq:heatengine_eta}, the Carnot bound depends explicitly on the Sharma--Mittal parameters through the temperature normalization.
	
	%%%%%%%%%%%%%%%%%%%%%%%%%%%%%%%%%%%%%%%%%%%%%%%%%%%%
	%%%%%%%%%%%%%%%%%%%%%%%%%%%%%%%%%%%%%%%%%%%%%%%%%%%%
	%\subsection{Effect of Massive Gravity Parameters}
	
\begin{figure}[htbp]
		\centering
		\includegraphics[width=0.45\linewidth]{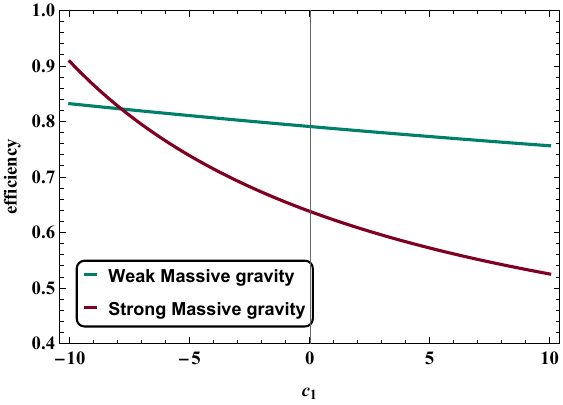}
		\caption{Representative dependence of the heat-engine efficiency on the massive-gravity coupling $c_1$ considering $q=1$, $R=0.05$, and $\delta=0.1$ for the weak ($m_g=0.4,\,C=0.1,\,c_2=0.2$) and strong ($m_g=0.8,\,C=0.1,\,c_2=23.0$) massive gravity regimes.}
		\label{fig:eta_massive_params}
	\end{figure}

	%%%%%%%%%%%%%%%%%%%%%%%%%%%%%%%%%%%%%%%%%%%%%%%%%%%%
	%%%%%%%%%%%%%%%%%%%%%%%%%%%%%%%%%%%%%%%%%%%%%%%%%%%%
	The massive gravity sector affects the heat input directly through the final two terms in Eq.~\eqref{eq:heatengine_QH}. The couplings $c_1$ and $c_2$ therefore modify the efficiency by changing the enthalpy difference along the upper isobar. In particular, increasing $c_1$ changes the contribution proportional to $r_2^2-r_1^2$, while increasing $c_2$ changes the contribution proportional to $r_2-r_1$. These effects are shown in Figure~\ref{fig:eta_massive_params}.
	
	%%%%%%%%%%%%%%%%%%%%%%%%%%%%%%%%%%%%%%%%%%%%%%%%%%%%
	%%%%%%%%%%%%%%%%%%%%%%%%%%%%%%%%%%%%%%%%%%%%%%%%%%%%
	%\subsection{Physical Interpretation}
	
	%%%%%%%%%%%%%%%%%%%%%%%%%%%%%%%%%%%%%%%%%%%%%%%%%%%%
	%%%%%%%%%%%%%%%%%%%%%%%%%%%%%%%%%%%%%%%%%%%%%%%%%%%%
	The heat-engine efficiency provides a complementary probe of the thermodynamic behaviour of the ModMax-dRGT black hole. While the criticality and Gibbs free-energy analyses characterize equilibrium phase transitions and global stability, the heat-engine construction quantifies the ability of the black hole to convert absorbed heat into useful work in the extended phase space.
	
	The ModMax parameter $\gamma$ enters through the effective charge combination
	\begin{equation}
		q_{\rm eff}^{\,2}=q^2e^{-\gamma},
	\end{equation}
	and therefore modifies the heat input and efficiency. Since $r_2>r_1$, the electromagnetic contribution in Eq.~\eqref{eq:heatengine_QH} is negative. Increasing $\gamma$ suppresses this term and changes the net heat absorbed along the upper isobar.
	
	The Sharma--Mittal parameters $(R,\delta)$ do not explicitly modify the rectangular-cycle efficiency when the cycle is fixed in the $P$-$V$ plane, because $W$ and $Q_H$ are determined by the enthalpy and volume. However, they modify the physical temperature assigned to each corner of the cycle and therefore affect the Carnot bound. This distinction shows that the ModMax sector primarily affects the enthalpy-based work-to-heat conversion, whereas the Sharma--Mittal sector modifies the thermal scale and limiting efficiency.
	
	The massive gravity parameters $(m_g,c_1,c_2)$ enter directly into the enthalpy and consequently influence both the heat input and the engine efficiency. Therefore, black hole heat engines provide an additional diagnostic tool for distinguishing the thermodynamic effects of nonlinear electrodynamics, massive gravity, and generalized entropy corrections.
	%%%%%%%%%%%%%%%%%%%%%%%%%%%%%%%%%%%%%%%%%%%%%%%%%%%%
	%%%%%%%%%%%%%%%%%%%%%%%%%%%%%%%%%%%%%%%%%%%%%%%%%%%%
	\section{Topological Analysis of Thermodynamic Potentials}\label{Sec5}
\begin{figure}[ht]
		\centerline{
			\includegraphics[width=0.45\linewidth]{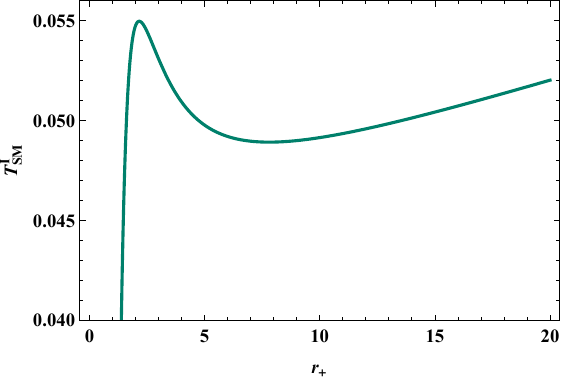} \hspace{0.2cm}
		}
		\caption{The behavior of $T_{\rm SM}^{\rm I}$ where $m_g=0.4,\,C=0.1,\,c_1=1,\,c_2=0.2,\,q=1,\,\gamma =0.5,\,\delta =0.1,\,R=0.05$.}\label{fig:THISM}
	\end{figure}

	%%%%%%%%%%%%%%%%%%%%%%%%%%%%%%%%%%%%%%%%%%%%%%%%%%%%
	%%%%%%%%%%%%%%%%%%%%%%%%%%%%%%%%%%%%%%%%%%%%%%%%%%%%
	To investigate the critical behavior of the Sharma--Mittal corrected temperature, we employ a topological approach. In this framework, the critical points of the property under study are treated as topological defects within a vector field, identified as zero points. By assigning a topological charge to these points, one can characterize the nature and stability of the black hole phase structure \cite{wei2022topology,alipour2023topological,yasir2024topological,hosseinifar2026study,zhang2023topology}.
	%Pressure
	%\begin{equation}
	%	\begin{aligned}
		%		P=&\frac{e^{-\gamma } \left(q^2 R-e^{\gamma } r_+^2 \left(C m_g^2 R (C c_2+c_1 r_+)-C c_1 \delta  m_g^2 r_++R\right)\right)}{4 r_+^2 \left(\pi  r_+^2 (2 R-3 \delta )-1\right)}\\&+\frac{e^{-\gamma } \left(e^{\gamma } r_+^2 \left(\pi  \delta  r_+^2-1\right) \left(C^2 c_2 m_g^2+1\right)+q^2 \left(\pi  \delta  r_+^2+3\right)\right)}{8 \pi  r_+^4 \left(\pi  r_+^2 (2 R-3 \delta )-1\right)}
		%	\end{aligned}
	%\end{equation}
	To this end, we define the isobaric Sharma--Mittal corrected temperature as
	\begin{equation}\label{TSM}
		T_{\rm SM}^{\rm I}=\frac{e^{-\gamma } \left(\pi  \delta  r_+^2+1\right)^{2-\frac{R}{\delta }} \left(e^{\gamma } r_+^2 \left(C m_g^2 (2 C c_2+c_1 r_+)+2\right)-4 q^2\right)}{4 \pi  r_+^3 \left(\pi  r_+^2 (3 \delta -2 R)+1\right)}.
	\end{equation}
	The behavior of the isobaric Sharma--Mittal corrected temperature for the weak massive gravity case, considering $C=0.1,\,m_g=0.4,\,c_1,\,c_2=0.2,\,q=1,\,\gamma =0.5,\,\delta =0.1,\,R=0.05,$ is shown in Figure \ref{fig:THISM}, which exhibits two critical points. 
	To analyze these points, we define a temperature-dependent potential as \cite{wei2022topology}
	\begin{eqnarray}
		\Phi =\frac{1}{\sin \theta} T_{\rm SM}^{\rm I},
	\end{eqnarray}
	This potential can be represented in the vector field via defining
	\begin{equation}\phi_r^\Phi=\partial_{r_+}\Phi,\;\;\phi_{\theta}^\Phi=\partial_{\theta}\Phi,
	\end{equation}
	which assuming $\phi=\phi_r+i\,\phi_\theta$, is normalized as $n^\Phi_r=\phi^\Phi_{r}/||\phi^\Phi||$ and $n^\Phi_{\theta}=\phi^\Phi_{\theta}/||\phi^\Phi||$.
	The vector field representation of the potential $\Phi$ for the specific choice of parameters $m_g=0.4,\,C=0.1,\,c_1=1,\,c_2=0.2,\,q=1,\,\gamma =0.5,\,\delta =0.1,\,R=0.05$ is depicted in Figure \ref{fig:THISM}. 
	As expected, two zero points appear in the vector field at coordinates $r_1=2.162$ and $r_2=7.804$, which are enclosed by the closed contours $\mathcal{C}_1$ and $\mathcal{C}_2$, respectively.
	To classify the nature of these points, we examine the deflection angle via the rotation of the vector field $\phi_r-\phi_\theta$ along the closed contours. A clockwise rotation signifies a topological charge of $-1$, whereas a counter-clockwise rotation indicates $+1$.
	In black hole thermodynamics, conventional zero points correspond to standard stable-unstable phase transition boundaries with topological charge of $-1$, where the thermodynamic potential exhibits a traditional extrema behavior. Conversely, novel zero points arise due to the non-trivial coupling with higher-order corrections and its topological charge is $+1$ \cite{wei2022topology}.
	\begin{figure}[ht]
		\centerline{
			\includegraphics[width=0.45\linewidth]{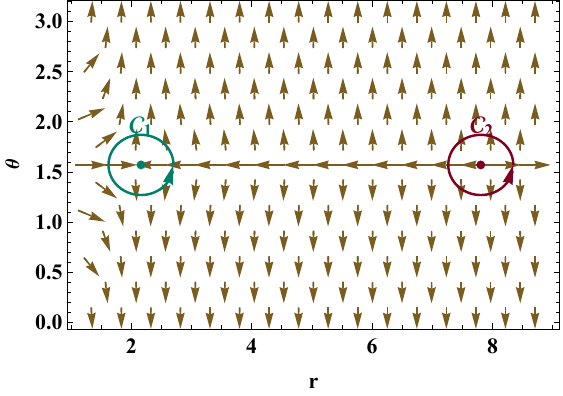} \hspace{0.2cm}
			\includegraphics[width=0.457\linewidth]{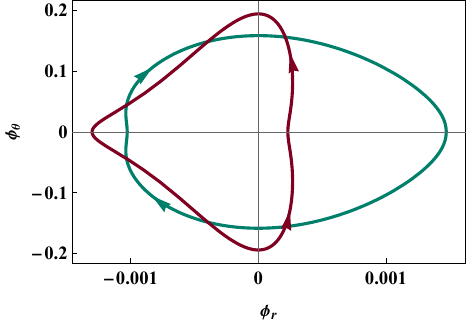} \hspace{0.2cm}
		}
		\caption{Left panel: Vector space of potential $\Phi$ where $m_g=0.4,\,C=0.1,\,c_1=1,\,c_2=0.2,\,q=1,\,\gamma =0.5,\,\delta =0.1,\,R=0.05$. Right panel: Topological charge identification via the rotation of $\phi_r-\phi_\theta$ curves around the zero points of $\Phi$.}\label{fig:topoTHISM}
	\end{figure}
	According to Figure \ref{fig:topoTHISM}, the topological charge associated with the zero point at $r_1$ is $-1$, and for the zero point at $r_2$, it is $+1$. Thus, we identified that $r_1$ is a conventional point and $r_2$ is a novel point. In conclusion, the topological study of the Sharma--Mittal corrected temperature reveals that the total topological charge of the system is conserved.

\begin{figure}[ht]
		\centerline{
			\includegraphics[width=0.45\linewidth]{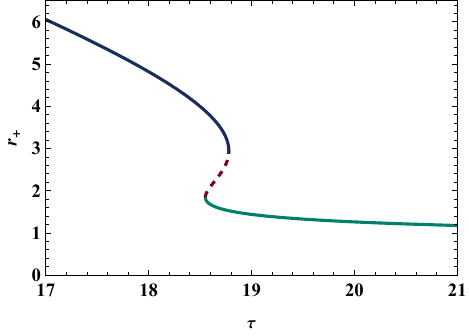} \hspace{0.2cm}
		}
		\caption{The behavior of $r-\tau$ where $P =0.0002,\, m_g=0.4,\,C=0.1;c_1=1,\,c_2=0.2,\,q=1,\,\gamma =0.5,\,\delta =0.1,\,R=0.05$. The curve displays three distinct phases based on the $r$ coordinate: small black hole $r < 1.846$, intermediate black hole $1.846 < r < 2.945$, and large black hole $r > 2.945$. Regarding the branching structure, the curve is characterized by a single branch outside the region $18.55 < \tau <18.78$, whereas a three-branch structure emerges within this interval.
%			The behavior of $r-\tau$ where $P =0.0002,\, m_g=0.4,\,C=0.1;c_1=1,\,c_2=0.2,\,q=1,\,\gamma =0.5,\,\delta =0.1,\,R=0.05$. This curve exhibits a single branch for $r < 1.846$ and $r > 2.945$, corresponding to small and large black hole phases, respectively. In the region $1.846 < r < 2.945$, the curve displays a three-branch structure, which characterizes the intermediate black hole phase.
			}\label{fig:tau}
%		
		
%		
%		The behavior of $r-\tau$ where $P =0.0002,\, m_g=0.4,\,C=0.1;c_1=1,\,c_2=0.2,\,q=1,\,\gamma =0.5,\,\delta =0.1,\,R=0.05$. This curve exhibits a single branch for $\tau < 18.55$ and $\tau > 18.78$. In the region $18.55 < \tau <18.78$, the curve displays a three-branch structure, which characterizes the intermediate black hole phase.
%		
%		The behavior of $r-\tau$ where $P =0.0002,\, m_g=0.4,\,C=0.1;c_1=1,\,c_2=0.2,\,q=1,\,\gamma =0.5,\,\delta =0.1,\,R=0.05$. This curve demonstrates three distinct regimes, in which $r < 1.846$ and $r > 2.945$, correspond to small and large black hole phases, respectively and in the region $1.846 < r < 2.945$, the curve characterizes the intermediate black hole phase.
%		
	\end{figure}

\begin{figure}[ht]
		\centerline{
			\includegraphics[width=0.45\linewidth]{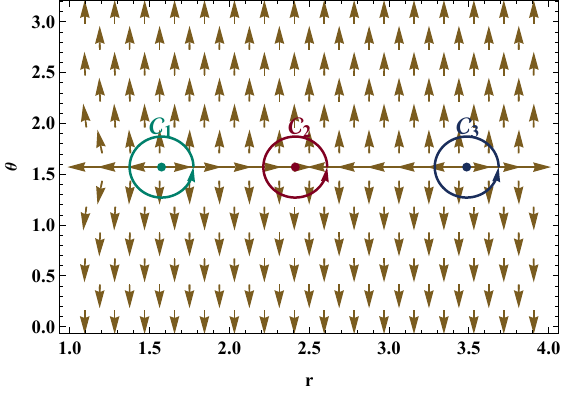} \hspace{0.2cm}
			\includegraphics[width=0.45\linewidth]{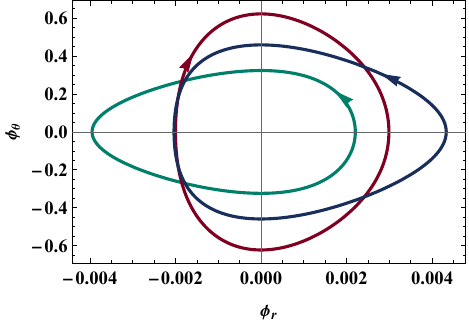} \hspace{0.2cm}
		}
		\caption{Left panel: Topological charge of potential $\mathcal{F}$ considering $P =0.0002,\, m_g=0.4,\,C=0.1;c_1=1,\,c_2=0.2,\,q=1,\,\gamma =0.5,\,\delta =0.1,\,R=0.05,\,\tau =18.70$. Right panel: The rotation of $\phi_r-\phi_\theta$ around the zero points. The direction of the deflection angle shows the topological charge of each critical point.}\label{fig:TopoGFE}
	\end{figure}

	To further study the thermodynamic stability and phase structure of the black hole, we extend our topological investigation to the off-shell generalized free energy of the black hole. This approach allows us to examine the stability of the black hole branches via the global topological properties of the free energy potential \cite{wu2024topological,gashti2025non,sadeghi2024thermodynamic,zare2026schwarzschild,sekhmani2025infrared,du2023topological}. We define the generalized free energy outside the horizon as
	\begin{equation}
		\begin{aligned}
			\mathcal{F}&=M-\frac{\mathcal{S}_{\rm SM}}{\tau}
			\\&
			=\frac{r_+}{12}  \left(6 C^2 c_2 m_g^2+3 C c_1 m_g^2 r_++16 \pi  P r_+^2+6\right)+\frac{e^{-\gamma } q^2}{2 r_+}+\frac{1-\left(\pi  \delta  r_+^2+1\right)^{R/\delta }}{R\, \tau }.
		\end{aligned}
	\end{equation}
	where $M$ and $\mathcal{S}$ represent the black hole mass and entropy, respectively, as defined in Eqs. \ref{eq:mass_therm_geom} and \eqref{eq:SSM}. Here, $\tau$ denotes the inverse of the black hole temperature outside the cavity. We represent the vector field associated with this potential as
	\begin{equation}\label{dF}
		\phi_r^\mathcal{F}=\partial_{r_+}\mathcal{F},\;
		\phi_{\theta}^\mathcal{F}=-\cot \theta \csc \theta.
	\end{equation}
	By setting $\phi_r^\mathcal{F} = 0$, we derive the relationship for $\tau$ as a function of the event horizon radius as
	\begin{equation}
		\tau=\frac{4 \pi  e^{\gamma } r_+^3 \left(\pi  \delta  r_+^2+1\right)^{\frac{R}{\delta }-1}}{e^{\gamma } r_+^2 \left(C^2 c_2 m_g^2+C c_1 m_g^2 r_++8 \pi  P r_+^2+1\right)-q^2}
	\end{equation}
	As shown in Figure \ref{fig:tau}, the behavior of the $r-\tau$ curve for the parameters $P =0.0002,\, m_g=0.4,\,C=0.1;c_1=1,\,c_2=0.2,\,q=1,\,\gamma =0.5,\,\delta =0.1,\,R=0.05$ exhibits two distinct turning points at the coordinates $(18.55,\,1.846)$ and $(18.78,\,2.945)$. 
	Within the interval $18.55 < \tau <18.78$, the curve displays three distinct branches, corresponding to small, intermediate, and large black hole phases. We restrict our analysis to this regime where the multi-branch structure is prevalent. Figure \ref{fig:TopoGFE} depicts the vector field representation of the potential $\mathcal{F}$ at $\tau =18.70$. 
	Three zero points are identified at coordinates $r_1=1.575$, $r_2=2.412$, and $r_3=3.486$, each enclosed by contours $\mathcal{C}_1$, $\mathcal{C}_2$, and $\mathcal{C}_3$.
	By examining the deflection angle of the vector field, we determine that the topological charge for the small and large black hole branches is $+1$, whereas the intermediate branch carries a charge of $-1$. Consequently, the total topological charge is $+1$. Furthermore, given that the innermost and outermost zero points both possess a charge of $+1$, we conclude that this black hole resides in the universal class $W^{1+}$ \cite{wei2024universal,chen2025universal}.
	%%%%%%%%%%%%%%%%%%%%%%%%%%%%%%%%%%%%%%%%%%%%%%%%%%%%
	%%%%%%%%%%%%%%%%%%%%%%%%%%%%%%%%%%%%%%%%%%%%%%%%%%%
	\section{Conclusion}\label{Sec6} 
	%%%%%%%%%%%%%%%%%%%%%%%%%%%%%%%%%%%%%%%%%%%%%%%%%%%%
	%%%%%%%%%%%%%%%%%%%%%%%%%%%%%%%%%%%%%%%%%%%%%%%%%%%%
	In this work, we studied the thermodynamic geometry, heat-engine efficiency, and topological phase structure of Sharma--Mittal corrected charged AdS black holes in ModMax nonlinear electrodynamics coupled to dRGT-like massive gravity. The results show that each sector of the theory has a distinct thermodynamic role. The ModMax parameter suppresses the effective charge contribution and shifts the microscopic interaction regime, the massive-gravity couplings modify the enthalpy and heat-engine performance, and the Sharma--Mittal parameters affect the corrected temperature and Carnot bound.
	
	The thermodynamic-geometry analysis shows that curvature singularities occur at the extremal boundary and at the degeneracy points of the thermodynamic metric. The sign of the thermodynamic Ricci scalar indicates a transition from charge-dominated attractive microscopic interactions at small horizon radius to massive-gravity-dominated repulsive behavior at larger radius. In particular, for a ModMax parameter equal to $0.8$, the effective charge contribution is reduced by about $55.1$ percent compared with the Maxwell case, weakening the attractive regime.
	
	The heat-engine analysis demonstrates that, for a fixed rectangular cycle in the pressure-volume plane, the efficiency is controlled by the enthalpy and thermodynamic volume. For the representative cycle considered here, the Maxwell-limit efficiency is approximately $0.5465$, while the efficiency decreases to approximately 0.5316 when the ModMax parameter is $0.8$. This corresponds to a relative reduction of about $2.73$ percent. The Sharma--Mittal parameters do not directly change the fixed-cycle mechanical efficiency, but they modify the corrected temperature and therefore the Carnot bound.
	
	The topological analysis reveals a rich phase structure. The corrected-temperature sector contains two critical points at horizon radii approximately $2.162$ and $7.804$, with topological charges minus one and plus one, respectively. Thus, the total topological charge in this sector is conserved and equal to zero. The generalized free-energy analysis shows three black hole branches at inverse temperature $18.70$, with zero points at horizon radii approximately $1.575$, $2.412$, and $3.486$. Their charges are plus one, minus one, and plus one, giving a total topological charge of plus one. Consequently, the black hole belongs to the universal thermodynamic topological class $W^{1+}$. Overall, the combined thermodynamic-geometry, heat-engine, and topological analyses provide a quantitative and consistent characterization of Sharma--Mittal corrected ModMax-dRGT black holes. ModMax nonlinearities reduce the effective charge sector and lower the engine efficiency, massive gravity modifies the enthalpy and microscopic interaction regime, and Sharma--Mittal entropy corrections reshape the thermal and topological properties of the system.
	
	%Furthermore, we employed a topological approach to characterize the critical behavior of thermodynamic properties of the black hole. Our analysis of the Sharma--Mittal corrected temperature demonstrated that critical points act as topological defects, which we classified into conventional and novel types based on their topological charges. Furthermore, by extending this topological framework to the generalized free energy perspective, we established a consistent classification of the phase structure of black hole. The identification of the total topological charge and the innermost/outermost zero points allowed us to classify the black hole into the $W^{1+}$ universal class.
	%%%%%%%%%%%%%%%%%%%%%%%%%%%%%%%%%%%%%%%%%%%%%%%%%%%%
	%%%%%%%%%%%%%%%%%%%%%%%%%%%%%%%%%%%%%%%%%%%%%%%%%%%%
	\hspace{4cm}
	\section*{Acknowledgements}
	H.H. is grateful to Excellence project FoS UHK 2203/2025-2026 for the financial support. Also, D.J.G. acknowledges the contribution of the COST Action CA21136  -- ``Addressing observational tensions in cosmology with systematics and fundamental physics (CosmoVerse)".
	%%%%%%%%%%%%%%%%%%%%%%%%%%%%%%%%%%%%%%%%%%%%%%%%%%%%
	%%%%%%%%%%%%%%%%%%%%%%%%%%%%%%%%%%%%%%%%%%%%%%%%%%%%

	\bibliographystyle{ieeetr}
%	\bibliography{ref.bib}

\end{document}